\documentclass{article}
\usepackage{ijcai22}

\usepackage{times}
\usepackage{soul}
\usepackage{url}
\usepackage[hidelinks]{hyperref}
\usepackage[utf8]{inputenc}
\usepackage[small]{caption}
\usepackage{graphicx}
\usepackage{amsmath}
\usepackage{amsthm}
\usepackage{booktabs}
\usepackage{algorithm}
\usepackage{multirow} 

\usepackage{blindtext}
\usepackage{algorithmic}
\usepackage{enumerate}
\usepackage{makecell}
\usepackage{graphicx}

\title{Image-text Retrieval: A Survey on Recent Research and Development}

\author{
Min Cao$^1$
\and
Shiping Li$^1$\and
Juntao Li$^1$\and
Liqiang Nie$^{2}$\And
Min Zhang$^{1,3}$
\affiliations
$^1$Soochow University,
$^2$ Shandong University,
$^3$ Harbin Institute of Technology, Shenzhen\\
\emails
\{mcao,ljt,minzhang\}@suda.edu.cn, spli@stu.suda.edu.cn,
nieliqiang@gmail.com}

\begin{document}

\maketitle

\begin{abstract}
In the past few years, cross-modal image-text retrieval (ITR) has experienced increased interest in the research community due to its excellent research value and broad real-world application.
It is designed for the scenarios where the queries are from one modality and the retrieval galleries from another modality.
This paper presents a comprehensive and up-to-date survey on the ITR approaches from four perspectives.
By dissecting an ITR system into two processes: feature extraction and feature alignment, we summarize the recent advance of the ITR approaches from these two perspectives.
On top of this, the efficiency-focused study on the ITR system is introduced as the third perspective.
To keep pace with the times, we also provide a pioneering overview of the cross-modal pre-training ITR approaches as the fourth perspective.
Finally, we outline the common benchmark datasets and evaluation metric for ITR, and conduct the accuracy comparison among the representative ITR approaches.
Some critical yet less studied issues are discussed at the end of the paper.
\end{abstract}

\section{Introduction}

Cross-modal image-text retrieval (ITR) is to retrieve the relevant samples from one modality as per the given user expressed in another modality, usually consisting of two sub-tasks: image-to-text (i2t) and text-to-image (t2i) retrieval.
ITR has extensive application prospects in the search field and is a valuable research topic.
Thanks to the prosperity of deep models for language and vision, we have witnessed the great success of ITR over the past few years~\cite{frome2013devise,li2021align}. 
For instance, along with the rising of BERT, the transformer-based cross-modal pre-training paradigm has gained momentum, and its pretrain-then-finetune form has been extended to the downstream ITR task, accelerating its development. 

 
\begin{figure}[t]
\begin{center}
   \includegraphics[width=1.0\linewidth, height=0.55\linewidth]{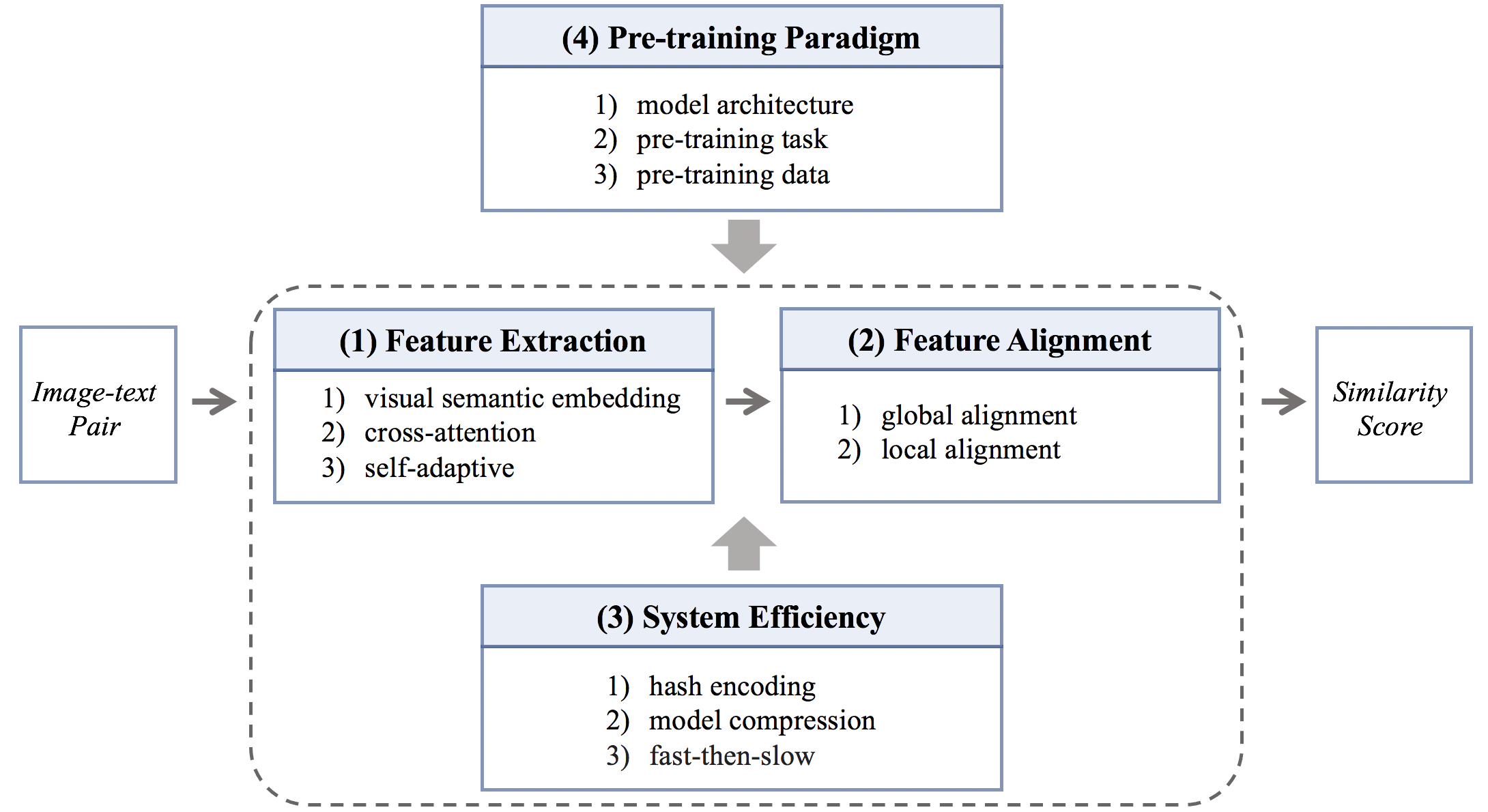}
\end{center}
   \caption{Illustration of the classification skeleton of ITR approaches from four perspectives.}
\vspace{-1em}
\label{fig1}
\end{figure}

It is worth mentioning that several prior efforts have been dedicated to conduct a survey on ITR. They, however, suffer from the following two limitations:
1) Beyond the ITR task, other multi-modal tasks, such as video-text retrieval and visual question answering, are also explored, resulting in a less in-depth ITR survey~\cite{uppal2022multimodal,baltruvsaitis2018multimodal};
2) the pre-training paradigm is largely untapped in existing surveys~\cite{chen2020review} and is indeed the mainstream nowadays.
In light of these, we conduct a comprehensive and up-to-date survey on the ITR task in this paper, especially with an investigation on the pre-training paradigm.

An ITR system generally consists of the feature extraction process with the image/text processing branches and the feature alignment process with an integration module. 
Contextualized in such an ITR system, we construct taxonomy from four perspectives to overview ITR approaches.
Figure~\ref{fig1} illustrates the classification skeleton for the ITR approaches.

\begin{enumerate}[(1)]
\item Feature extraction. 
Existing approaches on extracting robust and discriminative image and text features fall into three categories. 1) The visual semantic embedding based approaches work towards learning features independently. 2) The cross-attention approaches, by contrast, learn features interactively. And 3) the self-adaptive approaches aim at learning features with self-adaptive patterns. 
\item Feature alignment.
The heterogeneity of multimodal data makes the integration module important for aligning image and text features.
Existing approaches are in two variants. 
1) The global alignment-driven approaches align global features across modalities. 
2) Beyond that, some approaches attempt to find local alignment explicitly at a fine-grained level, so-called the local alignment-involved approaches. 

\item System efficiency. 
Efficiency plays a pivotal role in an excellent ITR system.
Apart from the research on improving ITR accuracy, a stream of works pursues a high-efficient retrieval system in three different ways.
1) The hash encoding approaches reduce the computational cost via binarizing the features in float format.
2) The model compression approaches emphasize low-energy and lightweight running. 
And 3) the fast-then-slow approaches perform the retrieval via a coarse-grained fast retrieval first and then a fine-grained slow one.  

\item Pre-training paradigm. 
To stand at the forefront of research development, we also gain insights into the cross-modal pre-training approaches for the ITR task, which has gained much attention recently.
Compared with the conventional ITR\footnote{We denote the ITR approaches without the benefit from cross-modal pre-training knowledge as the conventional ITR for distinguishing them from the ITR approaches under the cross-modal pre-training paradigm.}, the pre-training ITR approaches can benefit from the rich knowledge that is implicitly encoded by the large-scale cross-modal pre-trained models, yielding encouraging performance even without sophisticated retrieval engineering.
In the context of the ITR task, the cross-modal pre-training approaches are still applied to the taxonomy of the above three perspectives.
However, to characterize the pre-training ITR approaches more clearly, we re-classify them by three dimensions: model architecture, pre-training task, and pre-training data.
\end{enumerate}


In what follows, we summarize the ITR approaches based on the above taxonomy of the first three perspectives in Section~\ref{CITR}, and make a particular reference to the pre-training ITR approaches, i.e., the fourth perspective, in Section~\ref{ITRPP}.
A detailed overview of the common datasets, evaluation metric and accuracy comparison among representative approaches is presented in Section~\ref{DE}, followed by the conclusion and future work in Section~\ref{FW}. 






\section{Image-text Retrieval}
\label{CITR}






\subsection{Feature Extraction}
\label{FE}

\begin{figure}[t]
\begin{center}
   \includegraphics[width=1.0\linewidth, height=0.3\linewidth]{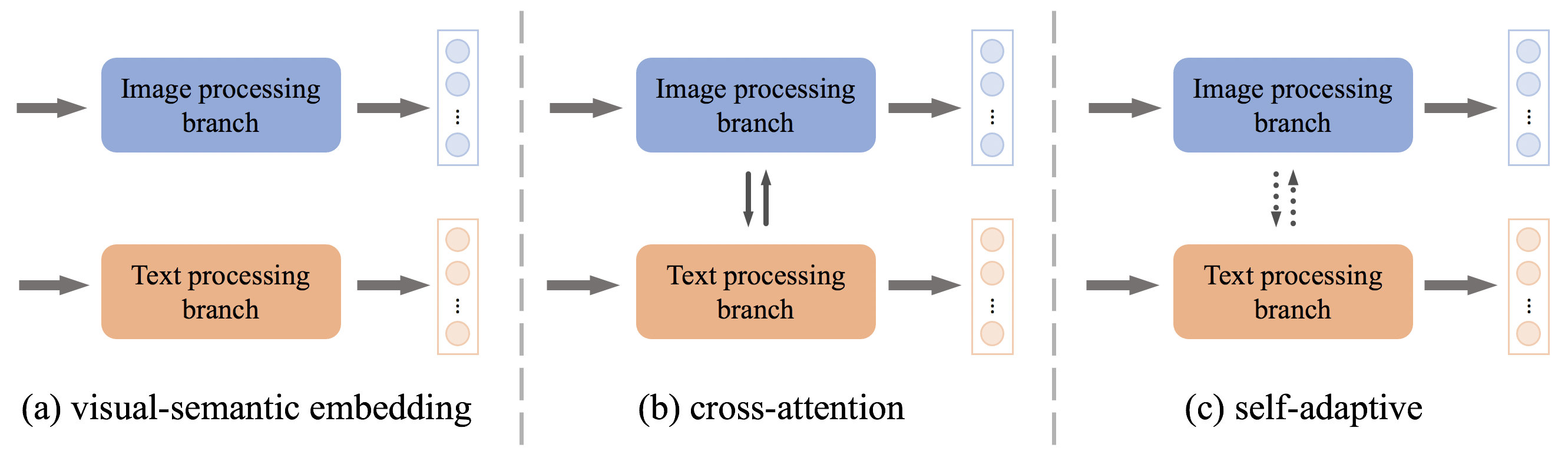}
\end{center}
   \caption{Illustration of different feature extraction architectures.}
\vspace{-1em}
\label{fig2}
\end{figure}

Extracting the image and text features is the first and also the most crucial process in the ITR system.
Under three different developing trends: visual semantic embedding, cross-attention and self-adaptive, as shown in Figure~\ref{fig2},
the feature extraction technology in ITR is thriving.



\textbf{Visual semantic embedding (VSE).}
Encoding image and text features independently is an intuitive and straightforward way for ITR, which was firstly proposed in~\cite{frome2013devise}.
Afterwards, such VSE based approaches are widely developed in roughly two aspects.
1) In terms of data, 
a stream of works~\cite{wu2019learning,chun2021probabilistic} tries to excavate the high-order data information for learning powerful features.
They learn features with equal treatment for all data pairs.
In contrast, some researchers~\cite{faghri2017vse++} propose to weight the informative pairs to improve the discrimination of features,
and others~\cite{hu2021learning} pay more attention to the mismatched noise correspondences in data pairs for the feature extraction.
Recently, riding the wave of the large-scale cross-modal pre-training technology, some works~\cite{jia2021scaling,huo2021wenlan} leverage large-scale web data directly to pre-train the image and text feature extractors, exhibiting the impressive performance on the downstream ITR task.
2) Regarding the loss function, the ranking loss is commonly used in the VSE based approaches~\cite{frome2013devise,faghri2017vse++} and constrains the inter-modal data relationship for learning features.
Besides that, \cite{wang2018learning} proposed a maximum-margin ranking loss with the neighborhood constraints for better extracting features.
\cite{zheng2020dual} proposed an instance loss explicitly considering the intra-modal data distribution.


Owing to the independent feature encoding, the VSE based approach enables a high-efficiency ITR system in which the features of massive gallery samples can be pre-computed offline.
However, it may bring suboptimal features and limited ITR performance due to less exploration of the interaction between the image and text data.

\textbf{Cross-attention (CA).}
\cite{lee2018stacked} made the first attempt to consider the dense pairwise cross-modal interaction and yielded tremendous accuracy improvements at the time.
Since then, various CA approaches have been put forward to extract features.
Employing the transformer architecture, the researchers can simply operate on a concatenation of image and text to the transformer architecture, thereby learning the cross-modal contextualized features.
It opens up a rich line of studies on the transformer-like CA approach~\cite{lu2019vilbert,chen2020uniter}.
Moreover, injecting some additional contents or operations into the cross-attention for assisting the feature extraction is also a new line of research.
\cite{ji2019saliency} adopted a visual saliency detection module to guide the cross-modal correlation.
\cite{cui2021rosita} integrated intra- and cross-modal knowledge to learn the image and text features jointly.



The CA approach narrows the data heterogeneity gap and tends to obtain high-accuracy retrieval results, yet comes at a prohibitive cost since each image-text pair must be fed into the cross-attention module online.

\textbf{Self-adaptive (SA).}
Instead of a fixed computation flow for extracting features in the VSE based and CA approaches,~\cite{qu2021dynamic} started from scratch and educed a self-adaptive modality interaction network in which different pairs can be adaptively inputted into different feature extraction mechanisms.
It powerfully inherits the respective merits of the above two groups and is classified as the SA approach. 


\subsection{Feature Alignment}
\label{FA}

After the feature extraction, it is desirable to align cross-modal features to compute pairwise similarity and achieve retrieval.
The global and local alignments are two directions.

\begin{figure}[t]
\begin{center}
   \includegraphics[width=1.0\linewidth, height=0.32\linewidth]{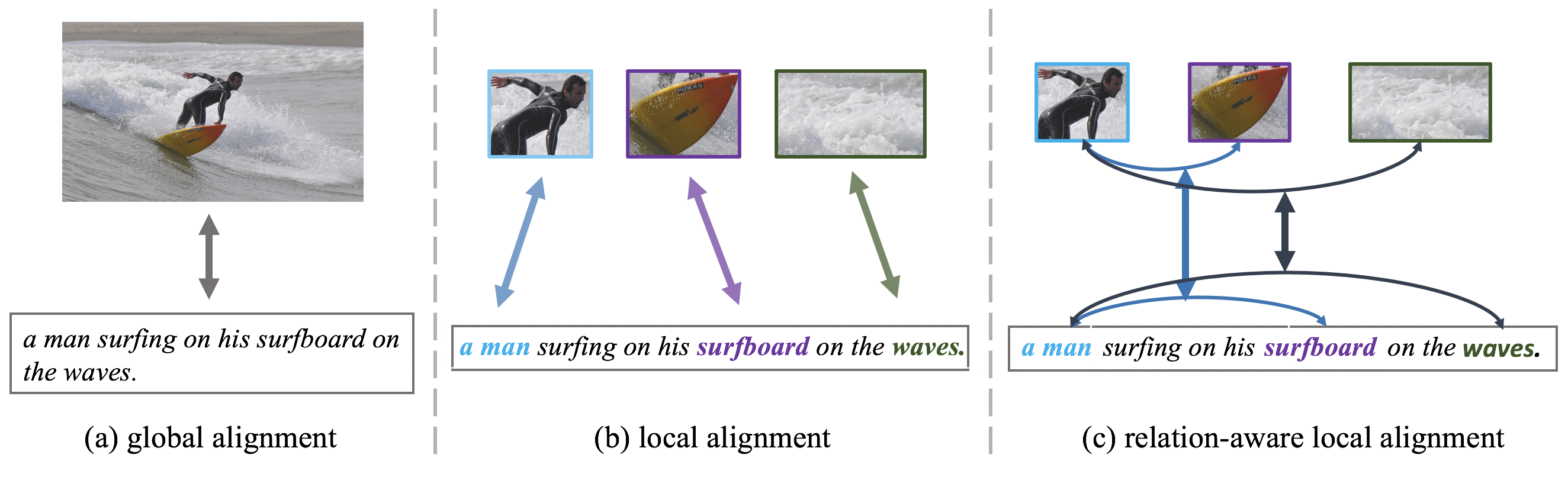}
\end{center}
   \caption{Illustration of different feature alignment architectures.}
\vspace{-1em}
\label{fig3}
\end{figure}

\textbf{Global alignment.}
In the global alignment-driven approach, the image and text are matched from a global viewpoint, as shown in Figure~\ref{fig3} (a).
Early works~\cite{faghri2017vse++,wang2018learning} are usually equipped with a clear and simple two-stream global feature learning network, and the pairwise similarity is computed by the comparison between global features.
Later studies~\cite{sarafianos2019adversarial,zheng2020dual} focus on improving such two-stream network architecture for better aligning global features.
Nonetheless, the above approaches with only the global alignment always present limited performance since the textual description usually contains finer-grained detail of image, which is prone to be smoothed by the global alignment.
However, there is an exception. 
The recent global alignment-driven approaches in a pretrain-then-finetune paradigm~\cite{jia2021scaling} tend to produce satisfactory results, attributed to the enlarging scale of pre-training data.

All in all, only applying the global alignment to ITR could lead to a deficiency of the fine-grained correspondence modeling and is relatively weak for computing reliable pairwise similarity.
Considering the alignment in other dimensions as a supplement to the global alignment is a solution.




\textbf{Local alignment.}
As shown in Figure~\ref{fig3} (b), the regions or patches within an image and words in a sentence correspond with each other, so-called the local alignment.
Global and local alignments form a complementary solution for ITR, which is a popular option and is classified as the local alignment-involved approach.
Adopting the vanilla attention mechanism~\cite{lee2018stacked,wang2019camp,chen2020uniter,kim2021vilt} is a trivial way to explore the semantic region/patch-word correspondences.
However, due to the semantic complexity, these approaches may not well catch the optimal fine-grained correspondences.
For one thing, attending to local components selectively is a solution for searching for an optimal local alignment.
\cite{liu2019focus} made the first attempt to align the local semantics across modalities selectively.
\cite{chen2020imram} and~\cite{zhang2020context} were not far behind.
The former learned to associate local components with an iterative local alignment scheme.
And the latter noticed that an object or a word might have different semantics under the different global contexts and proposed to adaptively select informative local components based on the global context for the local alignment.
After that, some approaches with the same goal as the above have been successively proposed with either designing an alignment guided masking strategy~\cite{zhuge2021kaleido} or developing an attention filtration technique~\cite{Diao2021SGRAF}.
For another thing, achieving the local correspondence in a comprehensive manner is also a pathway to approximate an optimal local alignment. 
\cite{wu2019unified} enabled different levels of textual components to align with regions of images.
\cite{ji2021step} proposed a step-wise hierarchical alignment network that achieves the local-to-local, global-to-local and global-to-global alignments.

Other than these, as shown in Figure~\ref{fig3} (c), there is another type of local alignment, \emph{i.e.}, the relation-aware local alignment that can promote fine-grained alignment. 
These approaches~\cite{xue2021probing,wei2020multi} explore the intra-modal relation for facilitating inter-modal alignment.  
In addition, some approaches~\cite{li2019visual,yu2020ernie,Diao2021SGRAF} model the image/text data as a graph structure with the edge conveying the relation information, and infer relation-aware similarities with both the local and global alignments by the graph convolutional network. 
Beyond these,~\cite{ren2021iais} considered the relation consistency, \emph{i.e.}, the consistency between the visual relation among the objects and the textual relation among the words.

\subsection{Retrieval Efficiency}

Combining the feature extraction in Section~\ref{FE} with the feature alignment in Section~\ref{FA} makes up a complete ITR system with attention to the retrieval accuracy.
Beyond that is the retrieval efficiency that is crucial for obtaining an excellent ITR system, thereby triggering a series of efficiency-focused ITR approaches.

\textbf{Hash encoding.}
The hash binary code introduces the advantage on the model's computation and storage, lightening a growing concern over the hash encoding approaches for ITR.
These studies learn to map the sample's feature to a compact hash code space to achieve a high-efficiency ITR.
\cite{yang2017pairwise} learned real-valued features and binary hash features of image and text simultaneously for benefiting from each other.
\cite{zhang2018attention} introduced an attention module to find the attended regions and words to prompt the binary feature learning.
Besides these approaches in a supervised setting, unsupervised cross-modal hashing is also a concern. 
\cite{li2018self} incorporated the adversarial network into the unsupervised cross-modal hashing to maximize the semantic correlation and consistency between two modalities.
\cite{yu2021deep} designed a graph-neighbor network to explore the sample's neighbor information for unsupervised hashing learning. 
The hash encoding approach benefits efficiency, yet also causes accuracy degradation due to the simplified feature representation with the binary code.

\textbf{Model compression.}
With the advent of the cross-modal pre-training age, ITR takes a giant leap forward in accuracy at the expense of efficiency.
The pre-training ITR approaches are usually characterized by the bulky network architecture,
which gives birth to the model compression approach.
Some researchers~\cite{gan2021playing} introduce the lottery ticket hypothesis to strive for smaller and lighter network architecture.
Moreover, based on the consensus, i.e., the image preprocessing process takes the most significant computing resource consumption in the pre-training architecture, some researchers~\cite{huang2020pixel,huang2021seeing} specifically optimize the image preprocessing process to improve the retrieval efficiency.
However, even with a lightweight architecture, most of these approaches that usually use the cross-attention for better feature learning still need to take a long reference time due to the quadratic executions of feature extraction.


\textbf{Fast-then-slow.}
The above two groups cannot bring the best compromise between efficiency and accuracy, raising the third group: the fast-then-slow approach.
Given that the VSE and CA approaches in Section~\ref{FE} have the efficiency and accuracy advantages, respectively, several researchers~\cite{miech2021thinking,li2021align} propose first to screen out large amounts of easy-negative galleries by the fast VSE technology and then retrieve the positive galleries by the slow CA technology, thereby striving for a good balance between efficiency and accuracy.





\section{Pre-training Image-text Retrieval}
\label{ITRPP}

For the ITR task, the early paradigm is to fine-tune the networks that have been pre-trained on the computer vision and natural language processing domains, respectively.
The turning point came in $2019$, and there was an explosion of interest in developing a universal cross-modal pre-training modal and extending it to the downstream ITR task~\cite{lu2019vilbert,li2019visualbert}. 
Under the powerful cross-modal pre-training technology,
the ITR task experiences explosive growth in performance without any bells and whistles.
Most of the pre-training ITR approaches currently adopt the transformer architecture as the building block.
On this foundation, the research mainly focuses on model architecture, pre-training task and pre-training data.

\textbf{Model architecture.}
A batch of works~\cite{lu2019vilbert,li2021align} is interested in the two-stream model architecture, \emph{i.e.}, two independent encodings followed by an optional later-interaction on the image and text processing branches.
Meanwhile, the one-stream architecture encapsulating the image and text processing branches into one gains popularity~\cite{li2019visualbert,li2020unicoder,kim2021vilt}.
Most approaches heavily rely on the image preprocessing process that usually involves an object detection module or convolution architecture for extracting the preliminary visual features and as the input of the follow-up transformer.
The resulting problems are twofold.
Firstly, this process consumes more computational resources than the subsequent processes, leading to the model's inefficiency.
Then, the predefined visual vocabulary from the object detection limits the model's expression ability, resulting in inferior accuracy.

Encouragingly, the research on improving the image preprocessing process has recently come into fashion.
Regarding improving efficiency, 
\cite{huang2021seeing} adopted a fast visual dictionary to learn the whole image's feature.
\cite{huang2020pixel} directly aligned the image pixels with the text in the transformer.
Alternatively, \cite{kim2021vilt,gao2020fashionbert} fed the patch-level features of the image into the transformer and \cite{liu2021kd} segmented the image into grids for aligning with the text.
In advancing accuracy, 
\cite{zhang2021vinvl} developed an improved object detection model to promote visual features.
\cite{xu2021e2e} put the tasks of object detection and image captioning together to enhance visual learning.
\cite{xue2021probing} explored the visual relation by adopting the self-attention mechanism when learning the image feature
Taking all these into account, \cite{dou2021empirical} investigated these model designs thoroughly and presented an end-to-end new transformer framework, reaching a win-win between efficiency and accuracy.
The advance of the cross-modal pre-training model architecture pushes forward the progress of ITR in performance.


\textbf{Pre-training task.}
The pre-training pretext task guides the model to learn effective multimodal features in an end-to-end fashion. 
The pre-training model is designed for multiple cross-modal downstream tasks, hence various pretext tasks are usually invoked.
These pretext tasks fall into two main categories: image-text matching and masked modeling.

The ITR is an important downstream task in the cross-modal pre-training domain and its associated pretext task, \emph{i.e.}, the image-text matching, is well received in the pre-training model.
In general, an ITR task-specific head is appended on the top of the transformer-like architecture to distinguish whether the input image-text pair is semantically matched by comparing the global features across modalities.
It can be viewed as an image-text coarse-grained matching pretext task~\cite{lu2019vilbert,li2020unicoder,chen2020uniter,li2021align,kim2021vilt}.
Furthermore, it is also expanded to the image-text fine-grained matching pretext tasks: patch-word alignment~\cite{kim2021vilt}, region-word alignment~\cite{chen2020uniter} and region-phrase alignment~\cite{liu2021kd}. 
There is no doubt that the pre-training image-text matching pretext task establishes a direct link to the downstream ITR task, which narrows the gap between the task-agnostic pre-training model and ITR.

Inspired by the pre-training in the natural language processing, the masked language modeling pretext task is commonly used in the cross-modal pre-training model.
Symmetrically, the masked vision modeling pretext task also emerges in this context.
Both are collectively called the masked modeling task.
In the masked language modeling task~\cite{lu2019vilbert,li2020unicoder,zhang2021vinvl}, the input text follows a specific masking rule that masks out several words in a sentence at random, and then this pretext task drives the network to predict the masked words based on the unmasked words and the input image.
In the masked vision modeling task, the network regresses the masked region's embedding feature~\cite{chen2020uniter} or predicts its semantic label~\cite{li2020unicoder} or does both~\cite{liu2021kd}.
The masked modeling tasks implicitly capture the dependencies between the image and text, providing powerful support to the downstream ITR task.


\textbf{Pre-training data.}
The research on the data level is an active trend in the cross-modal pre-training domain.
For one thing, the intra- and cross-modal knowledge in the image and text data are fully exploited in the pre-training ITR approaches~\cite{li2020oscar,cui2021rosita}.
For another, many studies concentrate on increasing the scale of pre-training data. 
Beyond the most widely used large-scale out-of-domain datasets, especially for the pre-training model~\cite{li2020unicoder,li2021align}, 
the in-domain datasets originally for fine-tuning and evaluating the downstream tasks are added into the pre-training data for better multimodal feature learning~\cite{li2020oscar,li2021align}.
Besides this,
the rich non-paired single-modal data can be added into the pre-training data for learning more generalizable features~\cite{li2020unimo}.
Other than all of these, some researchers~\cite{qi2020imagebert,jia2021scaling,yao2021filip} collect new larger-scale data for the pre-training model, and such a simple and crude operation usually brings outstanding performance on various downstream cross-modal tasks, including ITR.
In general, the focus at the data level positively affects the cross-modal pre-training model, naturally boosting the downstream ITR task.

\section{Datasets and Evaluation}
\label{DE}
\subsection{Datasets}
The researchers have proposed various datasets for ITR.
We summarize the most frequently used datasets as follows.
1) \textbf{COCO Captions} contains $123,287$ images collected from the Microsoft Common Objects in COntext (COCO) dataset, together with human generated five captions for each image.
The average length of captions is $8.7$ after a rare word removal. 
The dataset is split into $82,783$ training images, $5,000$ validation images and $5,000$ test images.
The researchers evaluate their models on the $5$ folds of 1K test images and the full 5K test images.
2) \textbf{Flickr30K} consists of $31,000$ images collected from the Flickr website.
Each image contains five textual descriptions. 
The dataset is divided into three parts, $1,000$ images for validation, $1,000$ images for the test, and the rest for training. 


\subsection{Evaluation Metric}
R@K is the most commonly used evaluation metric in ITR and is the abbreviation for recall at $K$-th in the ranking list, defined as the proportion of correct matchings in top-$K$ retrieved results. 

\subsection{Accuracy Comparison}

\begin{table}[tb!]
\begin{center}
\renewcommand\arraystretch{1.2}
\resizebox{0.47\textwidth}{!}{
\begin{tabular}{c|l|cc|cc|cc}
\Xhline{1.5pt}
\multicolumn{1}{c|}{\multirow{2}{*}{Type}}
&\multicolumn{1}{c|}{\multirow{2}{*}{Method}}& \multicolumn{2}{c|}{COCO1K} &\multicolumn{2}{c|}{COCO5K}&\multicolumn{2}{c}{Flickr30K}\\
& & \multicolumn{1}{c}{t2i}&\multicolumn{1}{c|}{i2t} & \multicolumn{1}{c}{t2i}&\multicolumn{1}{c|}{i2t} & \multicolumn{1}{c}{t2i}&\multicolumn{1}{c}{i2t}  \\
\hline
\hline
\multicolumn{1}{c|}{\multirow{5}{*}{VSE}}
 & VSE++~\cite{faghri2017vse++} & 52.0 & 64.6 & 30.3 & 41.3 & 39.6 & 52.9 \\
 & LTBNN~\cite{wang2018learning} & 43.3 & 54.9 & ~~- & ~~- & 31.7 & 43.2 \\ 
&TIMAM~\cite{sarafianos2019adversarial}  &~~-&~~-&~~-&~~-&42.6&53.1\\
&CVSE~\cite{wang2020consensus} &59.9&74.8&~~-&~~-&52.9&73.5\\
&PCME~\cite{chun2021probabilistic}    & {54.6}& {68.8}  &  31.9 & 44.2& ~~- & ~~-  \\ 
&ALIGN$^*$~\cite{jia2021scaling}&~~-&~~-&59.9&77.0&84.9&95.3\\ 
\hline
\multicolumn{1}{c|}{\multirow{7}{*}{CA}} 
&SCAN~\cite{lee2018stacked}&58.8&72.7&38.6&50.4&48.6&67.4\\ 
&SAN~\cite{ji2019saliency}&69.1&85.4&46.2&65.4&60.1&75.5\\ 
&Vilbert$^*$~\cite{lu2019vilbert}&~~-&~~-&~~-&~~-&58.2&~~-\\ 
&Oscar$^*$~\cite{li2020oscar}&78.2&89.8&57.5&73.5&~~-&~~-\\ 
&Uniter$^*$~\cite{chen2020uniter}&~~-&~~-&52.9&65.7&75.6&87.3\\ 
&Rosita$^*$~\cite{cui2021rosita}&~~-&~~-&54.4&71.3&74.1&88.9\\ 
&ALEBF$^*$~\cite{li2021align} &~~-&~~-&60.7&77.6&85.6&95.9\\ 
\hline
SA & DIME~\cite{qu2021dynamic}&64.8&78.8&43.1&59.3&63.6&81.0\\ 
\Xhline{1.5pt}
\end{tabular}}
\end{center}
\vspace{-1em}
\caption{Accuracy comparison at R@1 among the ITR approaches from the perspective of the feature extraction. The approach marked with `$*$' represents the pre-training approach. We show the best results of each approach reported in the original paper.}
\label{tab1}
\end{table}

\begin{table}[tb!]
\begin{center}
\renewcommand\arraystretch{1.2}
\resizebox{0.47\textwidth}{!}{
\begin{tabular}{c|l|cc|cc|cc}
\Xhline{1.5pt}
\multicolumn{1}{c|}{\multirow{2}{*}{Type}}
&\multicolumn{1}{c|}{\multirow{2}{*}{Method}}& \multicolumn{2}{c|}{COCO1K} &\multicolumn{2}{c|}{COCO5K}&\multicolumn{2}{c}{Flickr30K}\\
& & \multicolumn{1}{c}{t2i}&\multicolumn{1}{c|}{i2t} & \multicolumn{1}{c}{t2i}&\multicolumn{1}{c|}{i2t} & \multicolumn{1}{c}{t2i}&\multicolumn{1}{c}{i2t}  \\
\hline
\hline
\multicolumn{1}{c|}{\multirow{6}{*}{Global.}}

&{VSE++~\cite{faghri2017vse++} }    & {52.0}& {64.6}  &  30.3 & 41.3& 39.6& 52.9  \\ 
&LTBNN~\cite{wang2018learning} &43.3&54.9&~~-&~~-&31.7&43.2\\ 
&SEAM~\cite{wu2019learning}&57.8&71.2&~~-&~~-&52.4&69.1\\ 
&TIMAM~\cite{sarafianos2019adversarial}&~~-&~~-&~~-&~~-&42.6&53.1\\ 
&Dual-path~\cite{zheng2020dual}&47.1&65.6&25.3&41.2&39.1&55.6\\
&{PCME~\cite{chun2021probabilistic} }    & 54.6& 68.8  &  31.9 & 44.2& ~~- & ~~-  \\ 
&ALIGN$^*$~\cite{jia2021scaling}&~~-&~~-&59.9&77.0&84.9&95.3\\ 
\hline
\multicolumn{1}{c|}{\multirow{14}{*}{Local.}}
&SCAN~\cite{lee2018stacked}&58.8&72.7&38.6&50.4&48.6&67.4\\ 
&CAMP~\cite{wang2019camp} &58.5&72.3&39.0&50.1&51.5&68.1\\ 
&VSRN~\cite{li2019visual} &62.8&76.2&40.5&53.0&54.7&71.3\\ 
&IMRAM~\cite{chen2020imram}&61.7&76.7&39.7&53.7&53.9&74.1\\ 
&MMCA~\cite{wei2020multi}&61.6&74.8&38.7&54.0&54.8&74.2\\ 
&Uniter$^*$~\cite{chen2020uniter}&~~-&~~-&52.9&65.7&75.6&87.3\\ 
&SGRAF~\cite{Diao2021SGRAF}&63.2&79.6&41.9&57.8&58.5&77.8\\ 
&SHAN~\cite{ji2021step}&62.6&76.8&~~-&~~-&55.3&74.6\\ 
&ViLT$^*$~\cite{kim2021vilt}&~~-&~~-&42.7&61.5&64.4&83.5\\ 
&ALBEF$^*$~\cite{li2021align} &~~-&~~-&60.7&77.6&85.6&95.9\\
\Xhline{1.5pt}
\end{tabular}}
\caption{Accuracy comparison at R@1 among the ITR approaches from the perspective of the feature alignment. The approach marked with `$*$' represents the pre-training approach. Global. and Local. are short for the global-driven alignment and local-involved alignment approaches, respectively. We show the best results of each approach reported in the original paper.}
\vspace{-1em}
\label{tab2} 
\end{center}
\end{table}

We compare the representative and latest ITR approaches in terms of accuracy from two perspectives: feature extraction and feature alignment.

\textbf{Feature extraction.}
We present the comparison results in Table~\ref{tab1}.
For comparison among the VSE based approaches, ALIGN~\cite{jia2021scaling} achieves substantial improvement over others on accuracy thanks to the large-scale pre-training on more than one billion image-text pairs that far surpass the amount of data from other pre-training approaches.
For comparison among the CA approaches, we can see that the accuracy is improved gradually by these approaches over time.
For comparison between the VSE based and CA approaches, 
1) SCAN~\cite{lee2018stacked}, as the first attempt to the CA approach, makes a breakthrough at accuracy compared to the VSE based approach LTBNN~\cite{wang2018learning} at that time; 
2) taken as a whole, the CA approaches have the overwhelming advantage over the VSE based ones at R@1 aside from ALIGN, which is attributed to the in-depth exploration of cross-modal feature interaction in the CA approaches.
Nevertheless, as an exception, the VSE based approaches pre-training on extraordinarily massive data might offset the inferior performance caused by the less exploration on cross-modal interaction, which has been strongly supported by the results of ALIGN.
For comparison between the SA approaches and the VSE based and CA ones, under the same setting, \emph{i.e.}, conventional ITR, the SA approach DIME~\cite{qu2021dynamic} outperforms the VSE based and CA approaches on Flickr30k, and is inferior to the SAN~\cite{ji2019saliency} on COCO Captions.
There exists room for further development of the SA technology.

\textbf{Feature alignment.}
The comparison results are shown in Table~\ref{tab2}.
In terms of comparison within the global alignment-driven approaches, 
even with a basic two-stream architecture for the global alignment, ALIGN is still on top of other approaches at R@1, including TIMAM~\cite{sarafianos2019adversarial} and PCME~\cite{chun2021probabilistic} with the sophisticated network architecture for the global alignment.
In terms of comparison within the local alignment-involved approaches, ALBEF~\cite{li2021align} displays excellent performance. 
It is worth noting that Uniter~\cite{chen2020uniter} and ViLT~\cite{kim2021vilt} only with the vanilla attention mechanism can get decent results.
By contrast, SCAN~\cite{lee2018stacked} and CAMP~\cite{wang2019camp} with a similar mechanism are underwhelming at R@1.
The Uniter and ViLT conduct the ITR task in a pretrain-then-finetune form, and the rich knowledge from the pre-training cross-modal data benefits the downstream ITR task.
In terms of comparison between the global alignment-driven and local alignment-involved approaches, the latter shows better performance than the former on the whole, indicating the importance of local alignment for achieving high-accuracy ITR.

\begin{figure}[t]
\begin{center}
   \includegraphics[width=1\linewidth, height=0.4\linewidth]{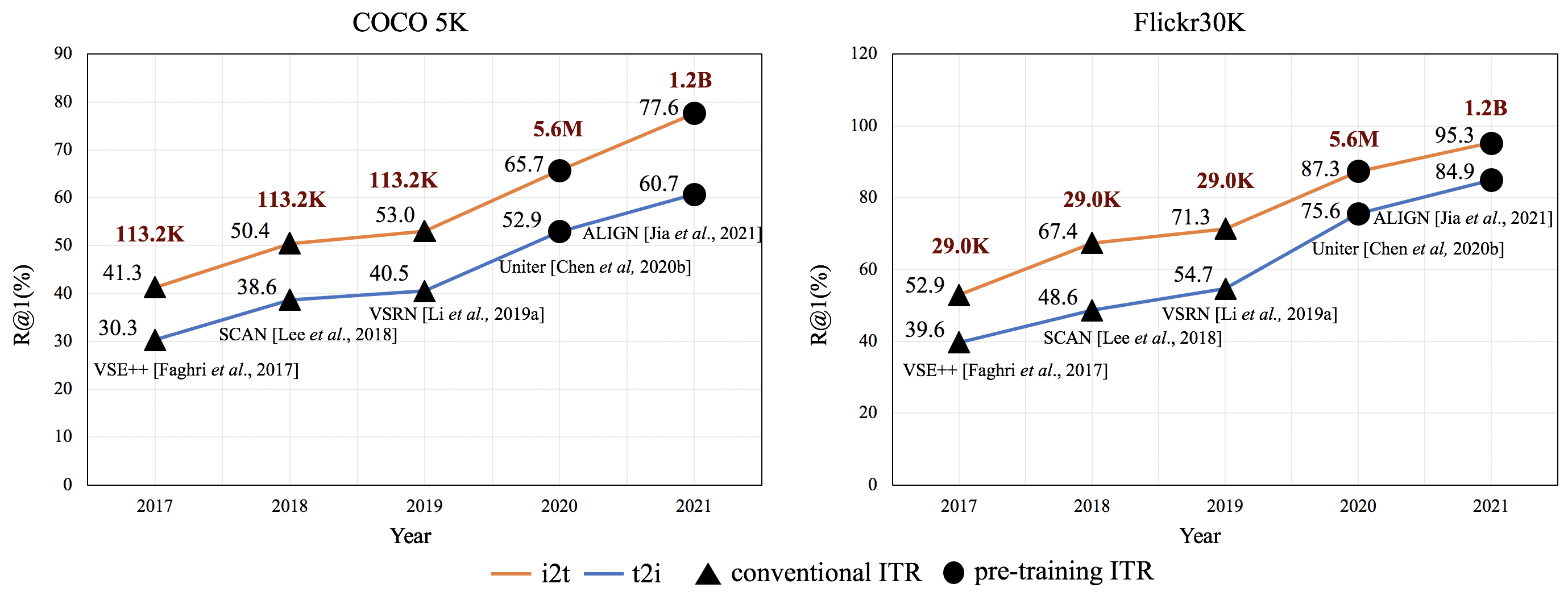}
\end{center}
   \caption{The development trend of ITR in recent years. The number with black above the line chart is the R@1 value of the approach, and the number with red is the amount of the multimodal training data.}
\vspace{-1em}
\label{fig4}
\end{figure}

Furthermore, we summarize the development trend of ITR from 2017 to 2021 in Figure~\ref{fig4}.
A clear trend of increasing accuracy can be seen over the years.
Specifically, the big jump comes in $2020$ thanks to the pre-training ITR technology.
After this, the accuracy of the pre-training ITR approach continues to keep the momentum of development. 
It follows that the pre-training ITR technology plays a leading role in promoting ITR development.
It can not be separated from the support of the enlarging scale of training data.
We can observe a dramatic increase in the amount of training data with the coming of the pre-training ITR.


\section{Conclusion and Future Works}
\label{FW}
In this paper, we presented a comprehensive review of ITR approaches from four perspectives: feature extraction, feature alignment, system efficiency and pre-training paradigm.
We also summarized extensively used datasets and evaluation metric in ITR, based on which we quantitatively analyzed the performance of the representative approaches.
It concludes that ITR technology has made considerable development over the past few years, especially with the coming of the cross-modal pre-training age.
However, there still exist some less-explored issues in ITR.
We make some interesting observations on possible future developments as follows.

\begin{figure}[t]
\begin{center}
   \includegraphics[width=0.82\linewidth, height=0.4\linewidth]{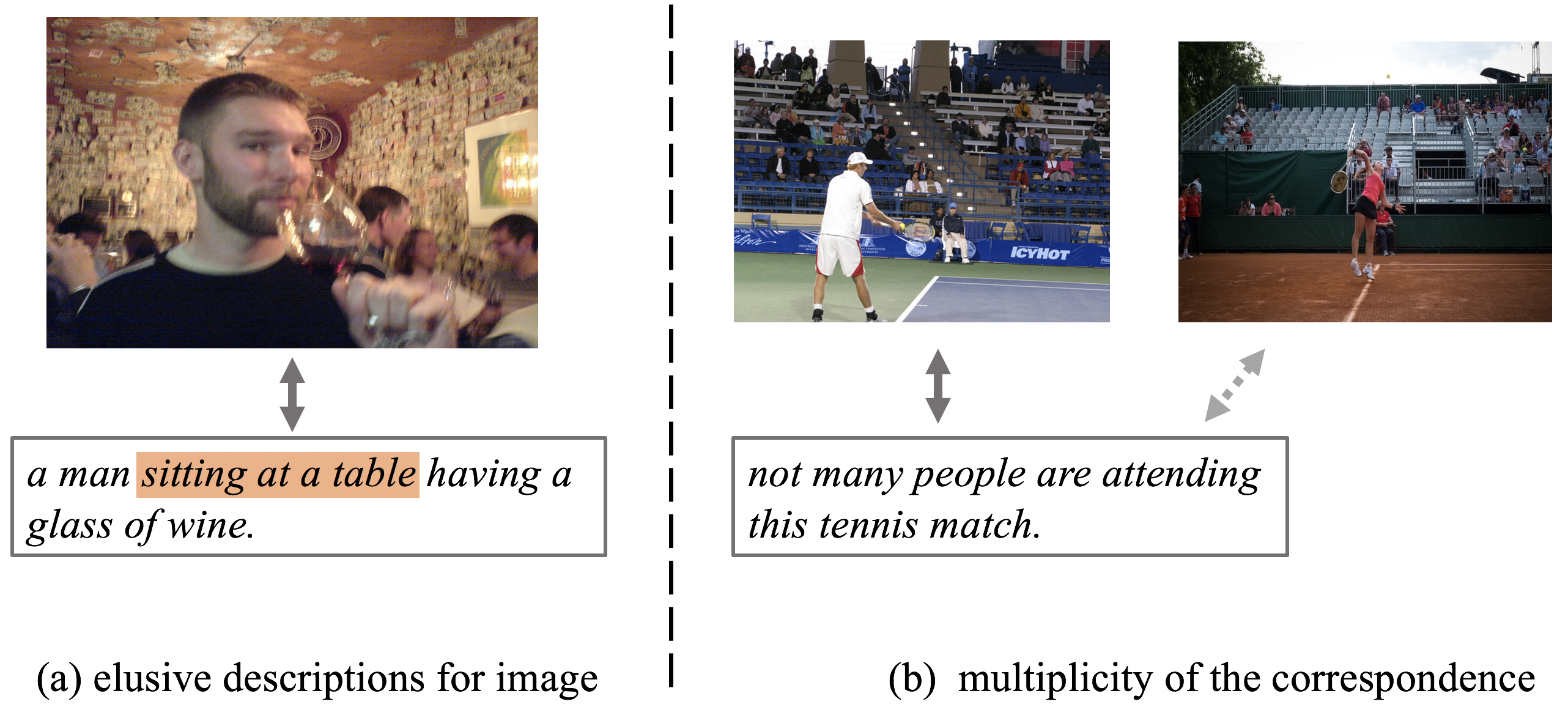}
\end{center}
   \caption{Illustration of noise data in the COCO Captions. (a) It is difficult to capture the content in the image based on the paired textual description highlighted by orange. (b) In addition to the positive image-text pair with a solid arrow, it seems to be correspondence for the negative image-text pair with a dotted arrow.}
\vspace{-1em}
\label{fig5}
\end{figure}

\textbf{Data.}
The current ITR approaches are essentially data-driven.
In other words, the researchers design and optimize the network for seeking an optimal retrieval solution based on available benchmark datasets.
For one thing, the heterogeneity and semantic ambiguity of cross-modal data can inevitably introduce noise into the datasets.
For example, as shown in Figure~\ref{fig5}, there exist the elusive textual description for the image and the multiplicity of the correspondences between the images and texts in the COCO Captions.
To some extent, therefore, the results of current ITR approaches on such datasets remain controversial.
There have been a few explorations about the data multiplicity~\cite{song2019polysemous,chun2021probabilistic,hu2021learning}, yet only considering the training data and ignoring the test one.
For another thing, beyond the vanilla data information, \emph{i.e.}, the image and text, the scene-text appearing in images is a valuable clue for ITR, which is usually ignored in the existing approaches.
\cite{mafla2021stacmr} is a pioneer work to explicitly incorporate the scene-text information into ITR model.
These studies leave room for further ITR development at the data level.


\textbf{Knowledge.}
Humans have the powerful ability to establish semantic connections between vision and language.
It benefits from their cumulative commonsense knowledge, together with the capacity of causal reasoning.
Naturally, incorporating this high-level knowledge into the ITR model is valuable for improving its performance.
CVSE~\cite{wang2020consensus} is a pioneer work that computes the statistical correlations in the image captioning corpus as the commonsense knowledge for ITR.
However, such commonsense knowledge is constrained by the corpus and is not a perfect fit for ITR.
It might be promising to tailor-make a commonsense knowledge and model the causal reasoning for ITR in the future.

\textbf{New paradigm.}
Under the current trend, the pre-training ITR approaches have an overwhelming advantage on accuracy compared to the conventional ITR ones.
The pretrain-then-finetune over a large-scale cross-modal model becomes a fundamental paradigm for achieving state-of-the-art retrieval results.
However, this paradigm with the need for large amounts of labeled data in the finetun phase is hard to apply in real-world scenarios.
It is meaningful to seek and develop a new resource-friendly ITR paradigm.
For example, the recently budding prompt-based tuning technology with an excellent few-shot capability provides a guide for developing such a new paradigm, so-called pretrain-then-prompt. 


\section{Acknowledgement}
This work is supported by the National Science Foundation of China under Grant NSFC 62002252, and is also partially supported by  Collaborative Innovation Center of Novel Software Technology and Industrialization. Juntao Li is the corresponding author of this paper.







\bibliographystyle{named}
\bibliography{ijcai22}

\end{document}